\documentclass{article}
\usepackage{spconf,amsmath,bm,graphicx}
\usepackage{url}
\usepackage[colorlinks,linkcolor=black,anchorcolor=black,citecolor=black,urlcolor=black]{hyperref}
\usepackage{amsfonts}

\title{Neural Speech Phase Prediction based on Parallel Estimation Architecture and Anti-Wrapping Losses}
\name{Yang Ai, Zhen-Hua Ling\thanks{This work was partially funded by the National Nature Science Foundation of China under Grant 61871358 and the Fundamental Research Funds for the Central Universities.}}
\address{National Engineering Research Center of Speech and Language Information Processing
\\University of Science and Technology of China, Hefei, P.R.China\\
{\small \tt \ yangai@ustc.edu.cn, zhling@ustc.edu.cn}}
\begin{document}
\ninept
\maketitle
\begin{abstract}
This paper presents a novel speech phase prediction model which predicts wrapped phase spectra directly from amplitude spectra by neural networks.
The proposed model is a cascade of a residual convolutional network and a parallel estimation architecture.
The parallel estimation architecture is composed of two parallel linear convolutional layers and a phase calculation formula, imitating the process of calculating the phase spectra from the real and imaginary parts of complex spectra and strictly restricting the predicted phase values to the principal value interval.
To avoid the error expansion issue caused by phase wrapping, we design anti-wrapping training losses defined between the predicted wrapped phase spectra and natural ones by activating the instantaneous phase error, group delay error and instantaneous angular frequency error using an anti-wrapping function.
Experimental results show that our proposed neural speech phase prediction model outperforms the iterative Griffin-Lim algorithm and other neural network-based method, in terms of both reconstructed speech quality and generation speed.

\end{abstract}
\begin{keywords}
speech phase prediction, parallel estimation architecture, anti-wrapping loss, neural network, phase wrapping
\end{keywords}
\section{Introduction}
\label{sec: Introduction}

Speech phase prediction, also known as speech phase reconstruction, recovers speech phase spectra from amplitude spectra and plays an important role in speech generation tasks.
Currently, several speech generation tasks, such as speech enhancement (SE) \cite{lu2013speech,xu2014regression,kim2020t}, bandwidth extension (BWE) \cite{wang2015speech,gu2016speech,li2015dnn} and speech synthesis (SS) \cite{zen2009statistical,takaki2017direct,wang2017tacotron,shen2018natural}, mainly focus on the prediction of amplitude spectra or amplitude-derived features (e.g., mel spectrograms and mel cepstra).
Therefore, speech phase prediction is crucial for waveform reconstruction in these tasks.
However, limited by the issue of phase wrapping and the difficulty of phase modeling, the precise prediction of the speech phase remains a challenge until now.

The Griffin-Lim algorithm \cite{griffin1984signal} is a well-known iterative phase estimation method which is widely used in several speech generation tasks.
However, the Griffin-Lim algorithm always causes unnatural artifacts in the reconstructed speech.
With the development of deep learning, Takamichi \MakeLowercase{\textit{et al.}} \cite{takamichi2018phase,takamichi2020phase} proposed a von-Mises-distribution deep neural network (DNN) for phase prediction.
However, the phase predicted by the DNN still needs to be refined using the Griffin-Lim algorithm.
Masuyama \MakeLowercase{\textit{et al.}} \cite{masuyama2020phase} proposed a DNN-based two-stage method which first predicted phase derivatives by DNNs, and then the phase was recursively estimated by a recurrent phase unwrapping algorithm.
To our knowledge, predicting speech wrapped phase spectra directly from amplitude spectra using neural networks has not yet been thoroughly investigated.

Due to the phase wrapping property, how to design 1) suitable architectures or activation functions to restrict the range of predicted phases for direct wrapped phase prediction and 2) loss functions suitable for phase characteristics, are the two major challenges for phase prediction based on neural networks.
To overcome these challenges, we propose a neural speech phase prediction model based on a parallel estimation architecture and anti-wrapping losses.
The proposed model passes the input log amplitude spectra through a residual convolutional network and a parallel estimation architecture to predict the wrapped phase spectra directly.
To restrict the output phase values to the principal value interval and predict the wrapped phases directly, the parallel estimation architecture imitates the process of calculating the phase spectra from the real and imaginary parts of complex spectra, and it is formed by two parallel convolutional layers and a phase calculation formula.
To avoid the error expansion issue caused by phase wrapping, we propose the instantaneous phase loss, group delay loss and instantaneous angular frequency loss activated by an anti-wrapping function at the training stage.
Experimental results show that our proposed model can achieve higher reconstructed speech quality than the iterative Griffin-Lim algorithm and the von-Mises-distribution DNN-based method.
In addition, our proposed model also exhibits the fastest generation speed, reaching 19.6x real-time on a CPU.

This paper is organized as follows.
In Section \ref{sec: Related works}, we briefly review the representative iterative speech phase estimation algorithm and neural network-based speech phase prediction method, respectively.
In Section \ref{sec: Proposed Methods}, we provide details on our proposed neural speech phase prediction model.
In Section \ref{sec: Experiments}, we present our experimental results.
Finally, we give conclusions in Section \ref{sec: Conclusion}.

\begin{figure*}
    \centering
    \includegraphics[height=2.6cm]{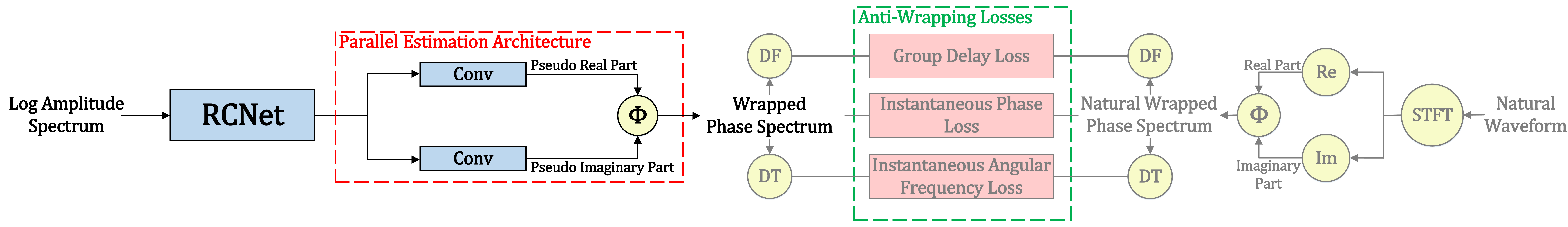}
    \caption{Details of the proposed neural speech phase prediction model. Here, \emph{RCNet}, \emph{Conv}, \emph{STFT}, \emph{DF}, \emph{DT}, \emph{Re}, \emph{Im} and \emph{$\Phi$} represent the residual convolutional network, linear convolutional layer, short-time Fourier transform, differential along frequency axis, differential along time axis, real part calculation, imaginary part calculation and phase calculation formula, respectively. Gray parts do not appear during generation.
    }
    \label{fig: Phase_model}
\end{figure*}

\section{Related work}
\label{sec: Related works}

\subsection{Iterative phase estimation}
\label{subsec: Iterative phase prediction}

This subsection briefly describes the well-known iterative Griffin-Lim algorithm \cite{griffin1984signal}.
It iteratively estimates the phase spectra from amplitude spectra via the short-time Fourier transform (STFT) and inverse STFT (ISTFT).
Assume that the amplitude spectrum is $\bm{A}\in \mathbb{R}^{F\times N}$, where $F$ and $N$ are the total number of frames and frequency bins, respectively.
Then initialize the phase spectrum $\hat{\bm{P}}\in \mathbb{R}^{F\times N}$ to zero matrix and iteratively execute the following formulas until convergence:
\begin{equation}
\label{equ: GL1}
\bm{S}=STFT\left[ISTFT\left(\bm{A}e^{j\hat{\bm{P}}}\right)\right],
\end{equation}
\begin{equation}
\label{equ: GL2}
e^{j\hat{\bm{P}}}=\bm{S}\oslash|\bm{S}|,
\end{equation}
where $\oslash$ and $|\cdot|$ represent the element-wise division and amplitude calculation, respectively.
The Griffin-Lim algorithm can be easily implemented and is popular in speech generation tasks.
Since the iterative algorithm always gives a local optimal solution, the reconstructed speech quality is limited by the influence of the initial phase and there are obvious artifacts in the reconstructed speech.

\subsection{Neural network-based phase prediction}
\label{subsec: Neural network-based phase prediction}

This subsection briefly describes the von-Mises-distribution DNN-based method \cite{takamichi2018phase,takamichi2020phase}.
This method assumes that the phase follows a von Mises distribution and then uses a DNN to predict the mean parameter of the phase distribution from the input log amplitude spectra at current and $\pm$2 frames.
The mean parameter is regarded as the predicted phase.
The DNN is composed of three 1024-unit feed-forward hidden layers activated by gated linear unit (GLU) \cite{dauphin2017language} and a linear output layer.
A multi-task learning strategy with phase loss and group delay loss is adopted to train the DNN.
The phase loss and group delay loss are formed by activating the phase error and group delay error using a negative cosine function, respectively.
Finally, the phase predicted by the DNN is set as the initial phase and refined by the Griffin-Lim algorithm with 100 iterations.

\section{Proposed Methods}
\label{sec: Proposed Methods}


\subsection{Model structure}
\label{subsec: Model Structure}

As shown in Figure \ref{fig: Phase_model}, the proposed neural speech phase prediction model predicts the wrapped phase spectrum $\hat{\bm{P}}\in \mathbb{R}^{F\times N}$ directly from the input log amplitude spectrum $\log\bm{A}\in \mathbb{R}^{F\times N}$ by a cascade of a residual convolutional network (RCNet) and a parallel estimation architecture.

In the RCNet, the input sequentially passes through a linear convolutional layer (kernel size=7 and channel size=512) and three parallel residual convolutional blocks (RCBlocks).
Then, the outputs of these three RCBlocks are summed (i.e., skip connections), averaged, and finally activated by a leaky rectified linear unit (LReLU) \cite{maas2013rectifier}.
Each RCBlock is formed by a cascade of three sub-RCBlocks.
In each sub-RCBlock, the input is first activated by an LReLU, then passes through a linear dilated convolutional layer, then is activated by an LReLU again, passes through a linear convolutional layer, and finally superimposes with the input (i.e., residual connections) to obtain the output.
The kernel sizes of all the convolutional operations in the three RCBlocks are 3, 7, and 11, respectively, and the channel sizes are 512.
The dilation factors of the dilated convolutional operations in the three sub-RCBlocks for each RCBlock are 1, 3, and 5, respectively.

The parallel estimation architecture is inspired by the process of calculating the phase spectra from the real and imaginary parts of complex spectra, and consists of two parallel linear convolutional layers (kernel size=7 and channel size=$N$) and a phase calculation formula $\bm{\Phi}$.
The outputs of the two parallel layers are the pseudo real part $\hat{\bm{R}}\in \mathbb{R}^{F\times N}$ and pseudo imaginary part $\hat{\bm{I}}\in \mathbb{R}^{F\times N}$, respectively.
Then the wrapped phase spectrum $\hat{\bm{P}}$ is calculated by $\bm{\Phi}$ as follows:
\begin{align}
\label{equ: Phase_calculate_matrix}
\hat{\bm{P}}=\bm{\Phi}(\hat{\bm{R}},\hat{\bm{I}}).
\end{align}
Equation \ref{equ: Phase_calculate_matrix} is calculated element-wise.
For $\forall R\in \mathbb{R}$ and $I\in \mathbb{R}$, we define
\begin{align}
\label{equ: Phase calculation}
\bm{\Phi}(R,I)=\arctan\left(\dfrac{I}{R}\right)-\dfrac{\pi}{2}\cdot Sgn^*(I)\cdot\left[Sgn^*(R)-1\right],
\end{align}
and $\bm{\Phi}(0,0)=0$.
When $x\ge 0$, $Sgn^*(x)$ is equal to $1$; otherwise, it is equal to $-1$.
Formula $\bm{\Phi}$ strictly restricts the predicted phase to the principal value interval $(-\pi,\pi]$ for direct wrapped phase prediction.

\subsection{Training criteria}
\label{subsec: Training Criteria}

\begin{figure}
    \centering
    \includegraphics[height=5cm]{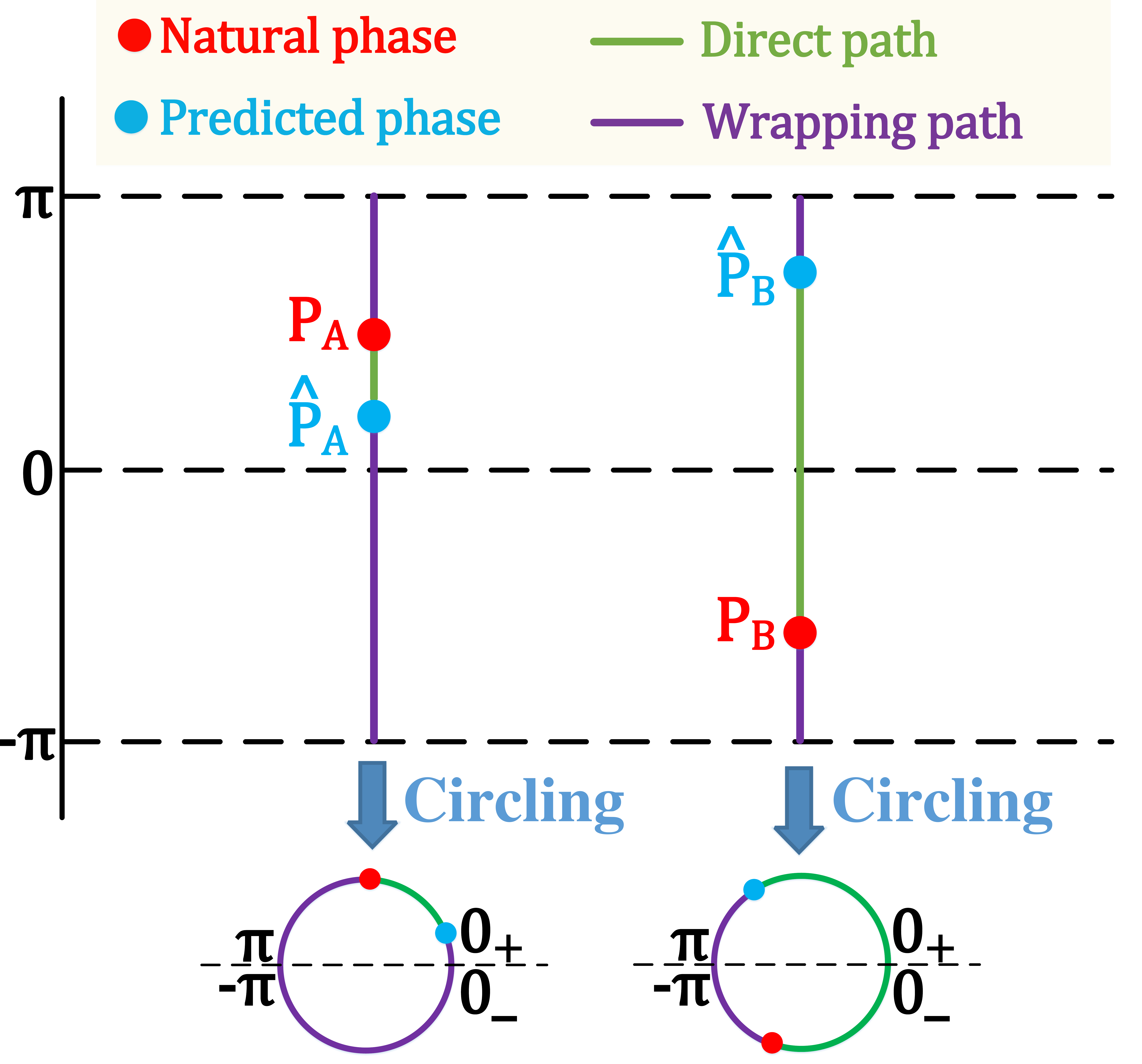}
    \caption{An illustration explanation of the error expansion issue caused by phase wrapping.
    }
    \label{fig: Phase_wrapping}
\end{figure}

Due to the wrapping property of the phase, the absolute error $e_a=|\hat{P}-P|$ between the predicted phase $\hat{P}$ and the natural phase $P$ might not be their true error.
As shown in Figure \ref{fig: Phase_wrapping}, assuming that the phase principal value interval is $(-\pi,\pi]$, there are two paths from the predicted phase point $\hat{P}_*$ to the natural one $P_*$, i.e., the direct path (corresponding to the absolute error) and the wrapping path (corresponding to the wrapping error).
Visually, we can connect the vertical line segment between $-\pi$ and $\pi$ end to end into a circle, according to the wrapping property of the phase.
Obviously, the wrapping path must pass through the boundary of the principal value interval, and the wrapping error is $e_w=2\pi-|\hat{P}-P|$.
Therefore, the true error between $\hat{P}$ and $P$ is
\begin{align}
\label{equ: True gap}
e=\min\{|\hat{P}-P|,2\pi-|\hat{P}-P|\}.
\end{align}
For example, in Figure \ref{fig: Phase_wrapping}, the true error between $\hat{P}_A$ and $P_A$ is the absolute error, but the true error between $\hat{P}_B$ and $P_B$ is the wrapping error.
This means that the absolute error and the true error satisfy $|\hat{P}-P|\ge e$, resulting in \emph{error expansion issue} when using the conventional L1 loss or mean square error (MSE) loss.
Equation \ref{equ: True gap} can be written in another form:
\begin{align}
\label{equ: True gap2}
e=\left| \hat{P}-P-2\pi\cdot round\left( \dfrac{\hat{P}-P}{2\pi} \right) \right|,
\end{align}
where $round$ represents the rounding.
Obviously, Equation \ref{equ: True gap2} is a function of error $\hat{P}-P$.
We define a function $f_{AW}(x)$ as follows:
\begin{align}
\label{equ: f_AW}
f_{AW}(x)=\left| x-2\pi\cdot round\left( \dfrac{x}{2\pi} \right) \right|,x\in\mathbb R.
\end{align}
$f_{AW}$ is an anti-wrapping function which can avoid the error expansion issue caused by phase wrapping because $f_{AW}(\hat{P}-P)=e$.

Specifically, we define the instantaneous phase loss $\mathcal L_{IP}$ between the wrapped phase spectrum $\hat{\bm{P}}$ predicted by our model and the natural wrapped phase spectrum $\bm{P}=\bm{\Phi}(\bm{R},\bm{I})$ as follows:
\begin{align}
\label{equ: Phase Loss}
\mathcal L_{IP}=\mathbb{E}_{\left(\hat{\bm{P}},\bm{P}\right)} \overline{f_{AW}\left(\hat{\bm{P}}-\bm{P} \right)},
\end{align}
where $f_{AW}(\bm{X})$ means element-wise anti-wrapping function calculation for matrix $\bm{X}$, and $\overline{\bm{Y}}$ means averaging all elements in the matrix $\bm{Y}$.
$\bm{R}$ and $\bm{I}$ are the real and imaginary parts of the complex spectrum extracted from the natural waveform.
To ensure the continuity of the predicted wrapped phase spectrum along the frequency and time axes, we also define the group delay loss $\mathcal L_{GD}$ and instantaneous angular frequency loss $\mathcal L_{IAF}$, which are both activated by the anti-wrapping function $f_{AW}$ to avoid the error expansion issue as follows:
\begin{align}
\label{equ: Group Delay Loss}
\mathcal L_{GD}=\mathbb{E}_{\left(\Delta_{DF}\hat{\bm{P}},\Delta_{DF}\bm{P}\right)} \overline{f_{AW}\left(\Delta_{DF}\hat{\bm{P}}-\Delta_{DF}\bm{P} \right)},
\end{align}
\begin{align}
\label{equ: PTD Loss}
\mathcal L_{IAF}=\mathbb{E}_{\left(\Delta_{DT}\hat{\bm{P}},\Delta_{DT}\bm{P}\right)} \overline{f_{AW}\left(\Delta_{DT}\hat{\bm{P}}-\Delta_{DT}\bm{P} \right)},
\end{align}
where $\Delta_{DF}$ and $\Delta_{DT}$ represent the differential along the frequency axis and time axis, respectively.
Finally, the training criteria of our proposed model is to minimize the final loss
\begin{align}
\label{equ: Total Loss}
\mathcal L=\mathcal L_{IP}+\mathcal L_{GD}+\mathcal L_{IAF}.
\end{align}

\section{Experiments}
\label{sec: Experiments}

\subsection{Data and feature configuration}
\label{subsec: Data and feature configuration}

A subset of the VCTK corpus \cite{veaux2016superseded} was adopted in our experiments.
We selected 11,572 utterances from 28 speakers and randomly divided them into a training set (11,012 utterances) and a validation set (560 utterances).
We then built the test set, which included 824 utterances from 2 unseen speakers (a male speaker and a female speaker).
The original waveforms were downsampled to 16 kHz for the experiments.
When extracting the amplitude spectra and phase spectra from natural waveforms, the window size was 20 ms, the window shift was 5 ms, and the FFT point number was 1024 (i.e., $N=513$).

\begin{table}
\centering
    \caption{Objective and subjective evaluation results among phase prediction methods. Here, ``$a\times$" represents $a\times$ real time.}
    \resizebox{8.8cm}{1.05cm}{
    \begin{tabular}{c | c c | c | c}
        \hline
        \hline
        & SNR(dB)$\uparrow$ & F0-RMSE(cent)$\downarrow$ & RTF$\downarrow$ & MOS$\uparrow$\\
        \hline
        \textbf{GT} & -- & -- & -- & 3.97$\pm$0.052 \\
        \hline
        \textbf{NSPP} & \textbf{8.26} & \textbf{10.0} & \textbf{0.051 (19.6$\times$)} & \textbf{3.95$\pm$0.055}\\
        \textbf{GL22} & 2.70 & 66.4 & 0.053 (18.9$\times$) & 2.92$\pm$0.10\\
        \textbf{GL100} & 3.35 & 32.5 & 0.23 (4.48$\times$) & 3.73$\pm$0.069\\
        \textbf{DNN+GL100} & 5.03 & 13.2 & 0.29 (3.45$\times$) & 3.86$\pm$0.057\\
        \hline
        \hline
    \end{tabular}}
\label{tab_results}
\end{table}

\subsection{Comparison among phase prediction methods}
\label{subsec: Comparison among phase prediction methods}

We first conducted objective and subjective experiments to compare the performance of our proposed neural speech phase prediction model and other phase prediction methods.
Note that the object for comparison here is the speech waveforms reconstructed from the amplitude spectra and the predicted phase spectra through ISTFT.
The descriptions of methods for comparison are as follows\footnote{Source codes are available at \url{https://github.com/yangai520/NSPP}. Examples of generated speech can be found at \url{https://yangai520.github.io/NSPP}.}:

\begin{itemize}
\item {}{\textbf{NSPP}}: The proposed neural speech phase prediction model.
The model details are given in Section \ref{sec: Proposed Methods}.
The model was trained using the AdamW optimizer \cite{loshchilov2018decoupled} with $\beta_1=0.8$ and $\beta_2=0.99$ on a single Nvidia 3090Ti GPU.
The learning rate decay was scheduled by a 0.999 factor in every epoch with an initial learning rate of 0.0002.
The batch size was 16, and the truncated waveform length was 8000 samples (i.e., 0.5 s) for each training step.
The model was trained until 3100 epochs.

\item {}{\textbf{GL$\bm{n}$}}: The iterative Griffin-Lim phase estimation algorithm \cite{griffin1984signal} mentioned in Section \ref{subsec: Iterative phase prediction} with $n$ iterations ($n=22$ and $n=100$ were used in the experiments).

\item {}{\textbf{DNN+GL100}}: The von-Mises-distribution DNN-based phase prediction method \cite{takamichi2018phase,takamichi2020phase} mentioned in Section \ref{subsec: Neural network-based phase prediction}.
    The phase spectra were first predicted by the DNN and then refined by the Griffin-Lim algorithm with 100 iterations.
    We reimplemented it ourselves.
    The training configuration of the DNN is the same as that of \textbf{NSPP}.

\end{itemize}

Two objective metrics used in our previous work \cite{ai2020neural} were adopted here to compare the reconstructed speech quality, including the signal-to-noise ratio (SNR), which was an overall measurement of the distortions of both amplitude and phase spectra, and root MSE of F0 (denoted by F0-RMSE), which reflected the distortion of F0.
To evaluate the generation efficiency, the real-time factor (RTF), which is defined as the ratio between the time consumed to generate all test sentences using a single Intel Xeon E5-2680 CPU core and the total duration of the test set, was also utilized as an objective metric.
Regarding the subjective evaluation, mean opinion score (MOS) tests were conducted to compare the naturalness of the speeches reconstructed by these methods.
In each MOS test, twenty test utterances reconstructed by these methods along with the natural utterances were evaluated by at least 30 native English listeners on the crowdsourcing platform of Amazon Mechanical Turk\footnote{\url{https://www.mturk.com}.} with anti-cheating considerations \cite{buchholz2011crowdsourcing}.
Listeners were asked to give a naturalness score between 1 and 5, and the score interval was 0.5.

\begin{figure}
    \centering
    \includegraphics[height=3.7cm]{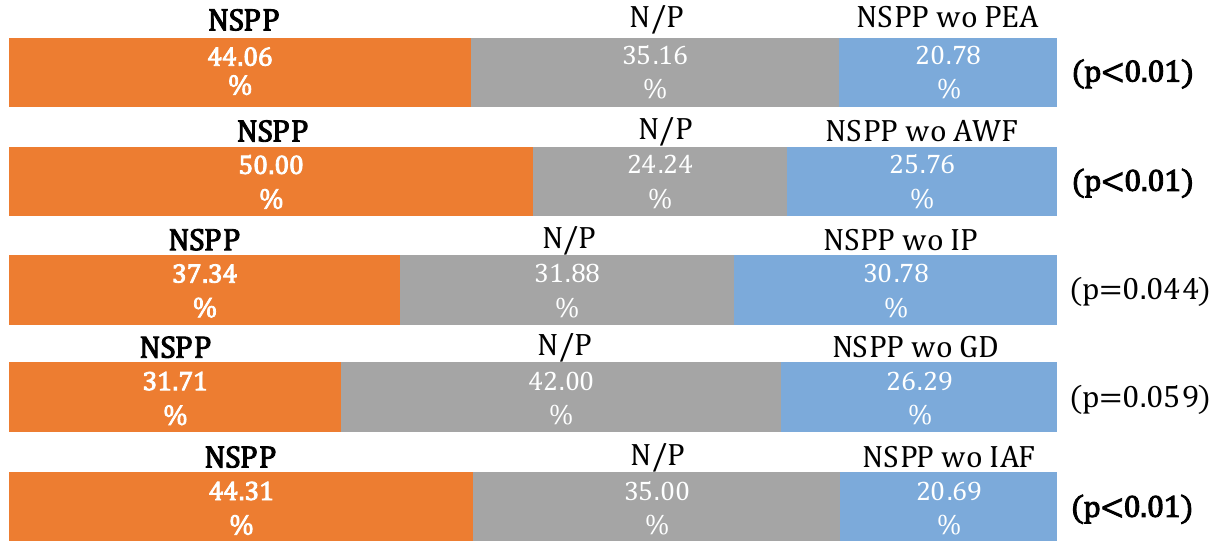}
    \caption{Average preference scores (\%) of ABX tests on speech quality between \textbf{NSPP} and its ablated variants, where N/P stands for ``no preference" and $p$ denotes the $p$-value of a $t$-test between two models.
    }
    \label{fig: ABX}
\end{figure}

Both the objective and subjective results are listed in Table \ref{tab_results}.
Our proposed \textbf{NSPP} obtained the highest SNR and the lowest F0-RMSE among all methods.
Regarding the subjective results of MOS scores, the \textbf{NSPP} also outperformed the other three methods significantly ($p<0.01$ of paired $t$-tests).
Besides, the MOS score of the \textbf{NSPP} also approached that of the ground truth natural speech (i.e., the \textbf{GT} in Table \ref{tab_results}), and the difference between the \textbf{NSPP} and \textbf{GT} was insignificant ($p=0.55$).
These results proved the precise phase prediction ability of our proposed model.
Regarding the RTF, our proposed \textbf{NSPP} was also an efficient model, reaching 19.6x real-time generation on a CPU.
At the same generation speed, the Griffin-Lim algorithm could only iterate 22 rounds (i.e., the \textbf{GL22}), and the reconstructed speech quality was far inferior to \textbf{NSPP}.
The \textbf{GL100}, although fully iterated, still performed worse than our proposed \textbf{NSPP} due to the audible unnatural artifact sounds.
Compared with the \textbf{GL100}, the performance of the \textbf{DNN+GL100} was significantly improved, which was consistent with the conclusion in the original paper \cite{takamichi2018phase,takamichi2020phase}.
However, our proposed \textbf{NSPP} outperformed the \textbf{DNN+GL100} in terms of both the reconstructed speech quality and generation speed.
Besides, the \textbf{NSPP} was a fully neural network-based method without the extra phase refinement operation, which can be easily implemented.
The \textbf{NSPP} was also proven to be universal, as it exhibited excellent performance on phase prediction for unseen speakers in the test set.
It is also worth mentioning that the training speed of the \textbf{NSPP} was also fast, with a training time of 27 hours on this dataset using a single Nvidia 3090Ti GPU.

\subsection{Ablation studies}
\label{subsec: Ablation studies}

We then conducted several ablation experiments to explore the roles of some key modules in our proposed \textbf{NSPP}.
The ablated variants of the \textbf{NSPP} for comparison included the following:
\begin{itemize}
\item {}{\textbf{NSPP wo PEA}}: Removing the parallel estimation architecture from the \textbf{NSPP}.
The output of the residual convolutional network passes through a linear layer without activation to predict the phase spectra, which is the same way as used in the von-Mises-distribution DNN-based method \cite{takamichi2018phase,takamichi2020phase}.

\item {}{\textbf{NSPP wo AWF}}: Removing the anti-wrapping function $f_{AW}$ from the \textbf{NSPP} and adopting L1 losses for $\mathcal L_{IP}$, $\mathcal L_{GD}$ and $\mathcal L_{IAF}$ at the training stage.

\item {}{\textbf{NSPP wo IP}}: Removing the instantaneous phase loss $\mathcal L_{IP}$ from the \textbf{NSPP} at the training stage.

\item {}{\textbf{NSPP wo GD}}: Removing the group delay loss $\mathcal L_{GD}$ from the \textbf{NSPP} at the training stage.

\item {}{\textbf{NSPP wo IAF}}: Removing the instantaneous angular frequency loss $\mathcal L_{IAF}$ from the \textbf{NSPP} at the training stage.

\end{itemize}

We conducted ABX preference tests on the Amazon Mechanical Turk platform to compare the differences between the \textbf{NSPP} and its ablated variants.
In each ABX test, twenty utterances were randomly selected from the test set reconstructed by two comparative models and evaluated by at least 30 native English listeners.
The listeners were asked to judge which utterance in each pair had better speech quality or whether there was no preference.
In addition to calculating the average preference scores, the $p$-value of a $t$-test was used to measure the significance of the difference between two models.

The ABX test results are shown in Figure \ref{fig: ABX}.
As expected, we can see that the \textbf{NSPP} outperformed the \textbf{NSPP wo PEA} significantly ($p<0.01$).
Specifically, the speech reconstructed by the \textbf{NSPP wo PEA} exhibited annoying loud noise similar to electric current, which significantly affected the sense of hearing due to the imprecise phase prediction.
One possible reason is that it was difficult for neural networks without the parallel estimation architecture to restrict the range of predicted phases, leading to failure of few anti-wrapping losses.
These results indicated that the parallel estimation architecture was essential to wrapped phase prediction.
The \textbf{NSPP} also outperformed the \textbf{NSPP wo AWF} significantly ($p<0.01$), which proved that the anti-wrapping function was helpful for avoiding the error expansion issue.
We also find that the high-frequency energy of the speech reconstructed by the \textbf{NSPP wo AWF} was completely suppressed, resulting in an extreme dull listening experience.
Interestingly, there were no obvious mispronunciation or F0 distortion in the speech reconstructed by the \textbf{NSPP wo AWF} (F0-RMSE=12.0 cent, comparable to that of \textbf{NSPP}).
For the three losses, removing $\mathcal L_{IAF}$ (i.e., \textbf{NSPP wo IAF}) led to a significant subjective performance degradation ($p<0.01$), manifested in the presence of obvious spectral horizontal stripes in the reconstructed speech, causing annoying loud noise.
When removing $\mathcal L_{IP}$ (i.e., \textbf{NSPP wo IP}) and $\mathcal L_{GD}$ (i.e., \textbf{NSPP wo GD}), the ABX test results show that the subjective differences were slightly insignificant ($p$ was slightly larger than 0.01).
However, we found that the reconstructed speech quality of the \textbf{NSPP wo IP} and \textbf{NSPP wo GD} indeed degraded, because the speech reconstructed by the \textbf{NSPP wo IP} exhibited few low-frequency spectrum corruption issues, resulting in F0 distortion and blurry pronunciation (F0-RMSE=21.2 cent, significantly higher than that of \textbf{NSPP}), and the \textbf{NSPP wo GD} attenuated the overall spectral energy of the reconstructed speech, resulting in a mild dull listening experience. 

\section{Conclusion}
\label{sec: Conclusion}

In this paper, we have proposed a novel neural speech phase prediction model, which utilizes a residual convolutional network along with a parallel estimation architecture to directly predict the wrapped phase spectra from input amplitude spectra.
The parallel estimation architecture is a key module which consists of two parallel linear convolutional layers and a phase calculation formula, strictly restricting the output phase values to the principal value interval.
The training criteria of the proposed model is to minimize a combination of the instantaneous phase loss, group delay loss and instantaneous angular frequency loss, which are all activated by an anti-wrapping function to avoid the error expansion issue caused by phase wrapping.
Experimental results show that the proposed model outperforms the iterative Griffin-Lim algorithm and the von-Mises-distribution DNN-based method, regarding the reconstructed speech quality.
Besides, the proposed model is easy to implement and exhibits a fast training speed and generation speed.
Ablation studies demonstrate that the parallel estimation architecture, anti-wrapping function and three losses are all useful.
Applying the neural speech phase prediction model to concrete speech generation tasks (e.g., SE, BWE and SS) will be the focus of our future work.

\bibliographystyle{IEEEbib}
\bibliography{strings,refs}

\end{document}